\documentclass{sig-alternate-05-2015}
\usepackage[utf8]{inputenc}
\usepackage{algorithm}
\usepackage{algorithmicx}
\usepackage{algpseudocode}
\usepackage{amsmath}
\usepackage{url}
\usepackage{hyperref}
\usepackage{subcaption}

\begin{document}

%\setcopyright{acmcopyright}
\conferenceinfo{CIKM Cup 2016}{USA}

\title{Cross-Device User Matching Based on Massive Browse Logs: The Runner-Up Solution for the 2016 CIKM Cup}

\numberofauthors{2}
\author{
\alignauthor Jianxun Lian \\
\affaddr{University of Science and Technology of China}\\
\affaddr{Hefei, China}
\email{jianxun.lian@outlook.com}
\alignauthor Xing Xie \\
\affaddr{Microsoft Research}\\
\affaddr{Beijing, China}
\email{xingx@microsoft.com}
}

\date{October 2016}

\maketitle

\begin{abstract}
As the number and variety of smart devices increase, users may use myriad devices in their daily lives and the online activities become highly fragmented. Building an accurate user identity becomes a difficult and important problem for advertising companies. The task for the CIKM Cup 2016 Track 1 was to find the same user cross multiple devices. This paper discusses our solution to the challenge. It is mainly comprised of three parts: comprehensive feature engineering, negative sampling, and model selection. For each part we describe our special steps and demonstrate how the performance is boosted. We took the second prize of the competition with an F1-score of 0.41669.
\end{abstract}

\keywords{User Linking; User Profiling; Cross-Device Behavior Analysis; CIKM Cup}

\captionsetup[table]{skip=2pt}
\captionsetup[figure]{skip=2pt}
\makeatletter
\newcommand{\thickhline}{%
    \noalign {\ifnum 0=`}\fi \hrule height 1pt
    \futurelet \reserved@a \@xhline
} 
\makeatother

\section{Introduction}
With the rapid development of smart devices, users now have myriad choices to connect to the Internet for daily activities. A user may do shopping with his/her smart phone, primary work on a laptop, and watch movies on a tablet. Unless a service supports persistent user identities (e.g. Facebook Login), the same user on different devices is viewed independently. It results in companies having to deal with weak user identities at device level. To perform sophisticated user profiling especially for online advertising, it is important to link the same users across multiple devices and integrate his/her digital traces together. 

At the Conference on Information and Knowledge Management (CIKM) 2016, the Data-Centric Alliance (DCA) provided a dataset for cross-device entity linking challenge\footnote{https://competitions.codalab.org/competitions/11171}. The dataset contained an anonymized browse log for a set of userIDs representing the same user across multiple devices. For each browse log, DCA provided the obfuscated site URL and HTML title. Some of the linked users were released as the training set. The participants need to identify the remaining matching user across multiple devices. Submissions were evaluated using F1 measure (a harmonic mean of precision and recall).

\begin{figure*}[htbp]
\centering
\includegraphics[width=0.8\textwidth]{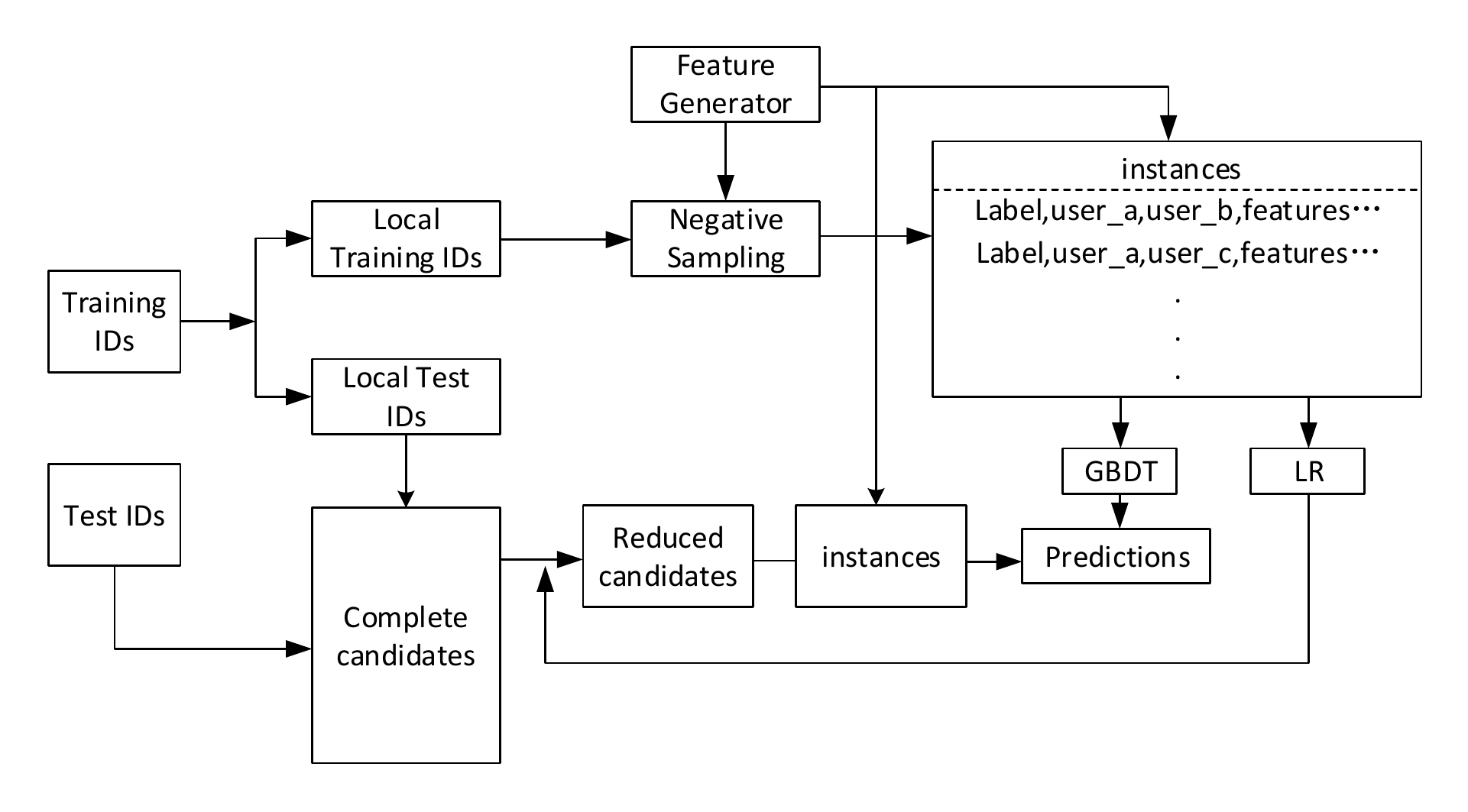}
\caption{The framework of our solution.}
\label{fig:pipeline}
\end{figure*}

In this paper, we describe our solution which placed 2nd at the competition. We formulated the task as a binary classification problem. Generally, it is simple, intuitive, and extremely effective. The three most essential parts are the feature engineering, negative sampling, and model selection. The framework is shown in Figure \ref{fig:pipeline}. Feature engineering is usually the most important factor for a data mining model. To achieve a satisfying score, we have designed comprehensive features from different levels. For the majority of data mining competitions, gradient boosted machine is the best single model and the ensemble of various models can further improve performance. We also consider this tip but conduct ensemble in a different way: we use the gradient boosted decision tree as the core classification model and use the logistic regression model to filter candidates. Since the complete candidate set is N $\times$ N which is too large in space, we have to do negative sampling. We find that the choice of negative instances significantly influences the performance of the model.

The remainder of this paper is organized as follows. In Section 2 we briefly review the data set. Then we describe our feature engineering approach in Section 3. In Section 4 and 5 we discuss our negative sampling algorithm and model selection, respectively. The online judging is presented in Section 6, followed by the conclusion in Section 7.

 \section{Dataset Overview}
 \begin{table}
     \centering
     \caption{Statistics of the provided dataset}
     \begin{tabular}{r|r} \hline
        \textbf{Statistics}  &  \textbf{Value} \\ \hline
        \#user  &    339,405 \\  
        \#fid   & \  \  \   14,148,535  \\  
        max.\#fid per user  & 2000 \\  
        min.\#fid per user  & 2 \\  
        avg.\#fid per user  & 196.8 \\  
        \#URL level-1  & 230,297 \\  
        \#URL level-2  & 1,435,418 \\  
        \#URL level-3  & 2,725,823 \\  
        \#URL level-4  & 4,644,424 \\  
        \#matched pairs for training & 506,136 \\  
        \#matched pairs to predict & 215,307 \\ \hline
     \end{tabular}
     \label{tab:statdataset} 
 \end{table}
There are four data files in total provided for the competition. The first one is facts.json which contains users' browsing logs. Each browsing log contains a list of events for a specific user, including fid(which can be regarded as an event ID), timestamp, and user id. All the IDs are anonymized. Another two files contains the mapping from an fid to the URL and HTML title, respectively. The last file offers a set of matching user IDs for training. The basic statistics of the dataset are showns in Table \ref{tab:statdataset}. We denote \textit{URL level}  as the length of the path, e.g., "bing.com" is level-1, "bing.com/images" is level-2.

\section{Feature Engineering}
We formulated the user-matching task as a binary classification problem. Since feature engineering is usually the most important part for data mining projects, we designed comprehensive features based on the browsing logs. Each instance is a pair of users with label 1 if the user pair is a match and label 0 if not. The feature set can be divided into three pillars as follows:
\subsection{General Similarity}
We assume that if two activity traces on two devices belong to the same user, then the traces will share some common websites or be similar in content. Thus, we design general similarity metrics from the perspective of words, events, URLs, and time: \\
\noindent \textbf{DocSim}: For each user we collect a bag of words from the title of HTML the user has visited. Based on this bag of words we calculate the words' weight in terms of TF-IDF and regard it as the user's document profile. For two users, their document similarity (DocSim) is measured as the cosine similarity between the document profiles:
\begin{equation}
    w^u_i = \frac{n_i}{\sum_{k}{n_k}} \times log \frac{|\{doc_j\}|}{|\{doc_j:n_i \in doc_j\}|}
\end{equation}
\begin{equation}
    \mathbf{w}^u= < w^u_0, w^u_1, ... , w^u_m>
\end{equation}
\begin{equation}
    DocSim(u,v)= \frac{\mathbf{w}^u \cdot \mathbf{w}^v}{\|\mathbf{w}^u\|\|\mathbf{w}^v\|}
\end{equation}
\\
\noindent \textbf{FidSim}: Similar to \textsl{DocSim}, however here we regard each \textsl{event ID} as a \textsl{word} and calculate the similarity based on the event document profiles. \\
\noindent \textbf{URLSim}: Similar to \textsl{DocSim}, however here we regard each \textsl{URL} as a \textsl{word} and calculate the similarity based on the URL document profiles. Since we consider 4 kinds of URL levels as shown in Table \ref{tab:statdataset}, we get 4 values from \textsl{URLSim}. \\
\noindent \textbf{FidComCnt,URLComCnt}: We count the number of common fid and URLs between the two users. \\
\noindent \textbf{HourCor}: We assume that users may have some temporal patterns in their online behaviors. For example, some users are active at midnight, while some users get up early in the morning. Thus we calculate the Pearson correlation coefficient based on the time distribution of two users:
\begin{equation}
    HourCor(u,v)=\frac{\sum_{i=1}^{24} (t^u_i - \overline{t^u})(t^v_i - \overline{t^v})}{\sqrt{\sum_{i=1}^{24}(t^u_i - \overline{t^u})^2}\sqrt{\sum_{i=1}^{24}(t^v_i - \overline{t^v})^2}}
\end{equation}
\\
\noindent \textbf{HourCE}: Consider the same concern with \textsl{HourCor}, but here we use cross entropy as the metrics:
\begin{equation}
    HourCE(u,v)=-\sum_{i=1}^{24} t^u_i \ log \ t^v_i
\end{equation}
\\
\noindent \textbf{DayCor}: Similar to \textsl{HourCor}, but here we calculate the Pearson correlation coefficient based on day distribution (from Monday to Sunday).\\
\noindent \textbf{DayCE}: Similar to \textsl{HourCE}, but here we calculate the cross entropy based on day distribution (from Monday to Sunday).
\\
\noindent \textbf{MonthCor,MonthCE}: Similar to \textsl{HourCor} and \textsl{HourCE} but from the point of month. \\
\noindent \textbf{FirstDateGap,LastDateGap}: The interval between the first/last dates of the two users, respectively. \\
\noindent \textbf{OverlapDay}: We count the number of dates both the two users are active.\\
\noindent \textbf{Skewness}: The ratio of shorter lifespan to longer lifespan:
\begin{equation}
    Skewness(u,v)=\frac{Min(lifespan(u),lifespan(v))}{Max(lifespan(u),lifespan(v))}
\end{equation}
\subsection{Key URLs}
The above feature pillar is called \textsl{General Similarity} because they are coarse-grained. For example, we calculate the number of mutual URLs between user $u$ and user $v$, but we don't know which particular URLs they are. Visiting \textsl{bing.com} is commonly happening among different users, while a common visit to a personal homepage strongly indicates a user matching. To this end, we plan to design more fine-grained features in this pillar which can describe what kind of URLs the two users share. We find that there are some URLs that appear more often in positive pairs than negative pairs. We assume that those URLs are key URLs that can differentiate matching users from dis-matching users. In order to find out those URLs,  for each URL $h$ we calculate the ratio of the probabilities that it appears simultaneously in a matching user pair to that in a random user pair:
\begin{equation}
    RatioLift(h) = \frac{\ \ \frac{|\{\# matching\ pairs\ containing\ h\}|}{|\{\# matching\ pairs\}|}\ \ }{\ \ \frac{|\{\# random\ pairs\ containing\ h\}|}{|\{\# random\ pairs\}|}\ \ }
\end{equation}
\begin{table}
    \centering
    \caption{Top 10 key URLs and their lift ratio.}
    \begin{tabular}{c|c} \thickhline
       URL (level-1)  &  RatioLift \\ \thickhline
       426ddb4efe252937/9db45ace43b3eb9c  & \ \ 5,686,956 \ \\ \hline
    449c90845cf62b1f/b82caf660250833b      &  \ \ 3,913,043\ \\ \hline
    449c90845cf62b1f/77cc413057b22ef2   &   \ \ 3,600,000 \ \\ \hline
     c0420384841e47d/16e720804d7385cb  &  \ \ 2,739,130\ \\ \hline
    449c90845cf62b1f/3cdf5b4cf0263a82   &  \ \ 2,647,826\ \\ \hline
     449c90845cf62b1f/1054834d358b06a2  &  \ \ 2,647,826\ \\ \hline
     09b0bf29d5bc1c1b/e0e89a73c6372042  &  \ \ 2,478,260\ \\ \hline
     5b67fb0f24569987/080473dc068d169c  & \ \ 2,269,565\  \\ \hline
     09b0bf29d5bc1c1b/ac4f7a44715b4762  &  \ \ 2,230,434\ \\ \hline
     967a94aa9df5ac93/16e720804d7385cb  &  \ \ 1,995,652\  \\ \thickhline
    \end{tabular}
    \label{tab:URLlift}
\end{table}
Table \ref{tab:URLlift} lists the top 10 key URLs. We can observe that these URLs are much more likely to appear in positive pairs than in negative pairs. There are about 5000 URLs with lift ratio above 2000. We categorize key URLs into 7 groups by the lift ratio: top 100, top 1000, top 2000, top 3000, top 4000, top 5000, and the others. For each user pair we count the number of key URLs in each group as features.
\subsection{Footprints}
We plan to design further fine-grained features in this pillar. We want to include the detailed activities of the users and meanwhile avoid overfitting.\\
\noindent \textbf{KeyURLDist}: We sort the key URLs by their lift ratio and divide the top 4000 key URLs into 40 buckets, with each bucket containing 100 URLs. For each user-user pair, we count the number of their common URLs in each bucket respectively. \\
\noindent \textbf{TopURLHit}: Since the top URLs shows an extremely high probability of matching users, in this feature, we use a 500-dimension indicator vector to record whether the top 500 key URLs exist in the common space of the two users. \\
\noindent \textbf{TemporalDist}: In the \textsl{General Similarity} pillar, we calculate the Pearson correlation similarity and cross entropy between two users' temporal (hour/day/month) distributions. Here we use the original temporal distributions as features. For example, for features in hour granularity, we use a 24-dimension vector to record the hourly activity amount. 

\floatname{algorithm}{Algorithm}
\renewcommand{\algorithmicrequire}{\textbf{Input:}}
\renewcommand{\algorithmicensure}{\textbf{Output:}}

\begin{algorithm}  
    \caption{Iterative Negative Sampling}
    \label{alg:ns}
    \begin{algorithmic}
        \Require $\mathbf{U}$, $\mathbf{M}$, $n$ and $k$
        \Ensure $\mathbf{S}$  
        \State
        \State $model \gets $ NULL
        \For{$i=0 \to k-1$}
            \State $\mathbf{S} \gets \varnothing $
            \For{$u \in \mathbf{U}$}
                \If{$model = $ NULL}
                    \State Randomly sample $n$ users from $\mathbf{U}$ and add \State the $n$ pairs to $\mathbf{S}$
                \Else
                    \State Select top $\frac{n}{2}$ users from $\mathbf{U}$ according to the 
                    \State $model$ and add to $\mathbf{S}$
                    \State Randomly sample $\frac{n}{2}$ users from $\mathbf{U}$ and add to
                    \State $\mathbf{S}$
                \EndIf
            \EndFor
            \State $\mathbf{S} \gets \mathbf{S} \cup \mathbf{M}$
            \State re-train $model$ based on $\mathbf{S}$
        \EndFor
        \State \Return{$\mathbf{S}$}
    \end{algorithmic}
\end{algorithm}

\subsection{Feature Evaluation}
We reserve 1000 users from the training file as our local validation set\footnote{A common practice is that we need to generate k validation sets in order to perform significance test on experimental results. Due to time limit we just skip this step.}. There are 3076 matching pairs in the local validation set. Table \ref{tab:featureperf} shows how the performance is improved when we add more fine-grained features. Adding \textsl{footprints} features significantly improves all the evaluation metrics, which demonstrate that fine-grained features play an essential role in user profiling. Table \ref{tab:topfeatures} lists the top 10 most important features according to the build-in feature evaluation functionality of gradient boosting machine \cite{friedman2001greedy}. It further demonstrates that the footprint features carry the most discriminative information.
\begin{table}
    \centering
    \caption{Performance evaluation with feature incrementation. Row \textsl{General-Sim} means using feature pillar 1 only. \textsl{+Key URLs} mean using feature \textsl{General-Sim} and \textsl{Key URLs}. Finally \textsl{+Footprints} means use all the three feature pillars.}
    \begin{tabular}{c|c|c|c|c} \hline \hline
  features  & AUC  & Recall  & Precision  & F1  \\ \hline
  General-Sim &\  0.8786  \  & \  0.4029  \   &\  0.4958   \  &\  0.4445   \    \\ \hline
  +Key URLs  & 0.8810  & 0.4091  & 0.5034  & 0.4513  \\ \hline
  +Footprints  & \textbf{0.9383}  & \textbf{0.5613}  & \textbf{0.6906}  & \textbf{0.6193}  \\ \hline \hline
    \end{tabular} 
    \label{tab:featureperf}
\end{table}

\begin{table}
    \centering
    \caption{Top 10 most important features.}
    \begin{tabular}{|c|c|} \hline  
\ \   \textbf{Feature Name}\ \  &\ \  \textbf{Split Gain}\ \  \\ \hline
    KeyURLDist02  & 1.0 \\ \hline
    HourCorrelation  & 0.4499 \\ \hline
    FidSim  & 0.4246 \\ \hline
    KeyURLDist01  & 0.4221 \\ \hline
    URLSim Level-1  & 0.3478 \\ \hline
    TopURLHit10  &  0.3183\\ \hline
    KeyURLDist00  & 0.3137 \\ \hline
    OverlapDay  & 0.2910 \\ \hline
    KeyURLDist07  & 0.2494 \\ \hline
    KeyURLDist08  & 0.2389 \\ \hline
    \end{tabular} 
    \label{tab:topfeatures}
\end{table}

\begin{figure*}[htbp]
\centering
\begin{subfigure}{.24\textwidth}
  \centering
  \includegraphics[width=1\textwidth]{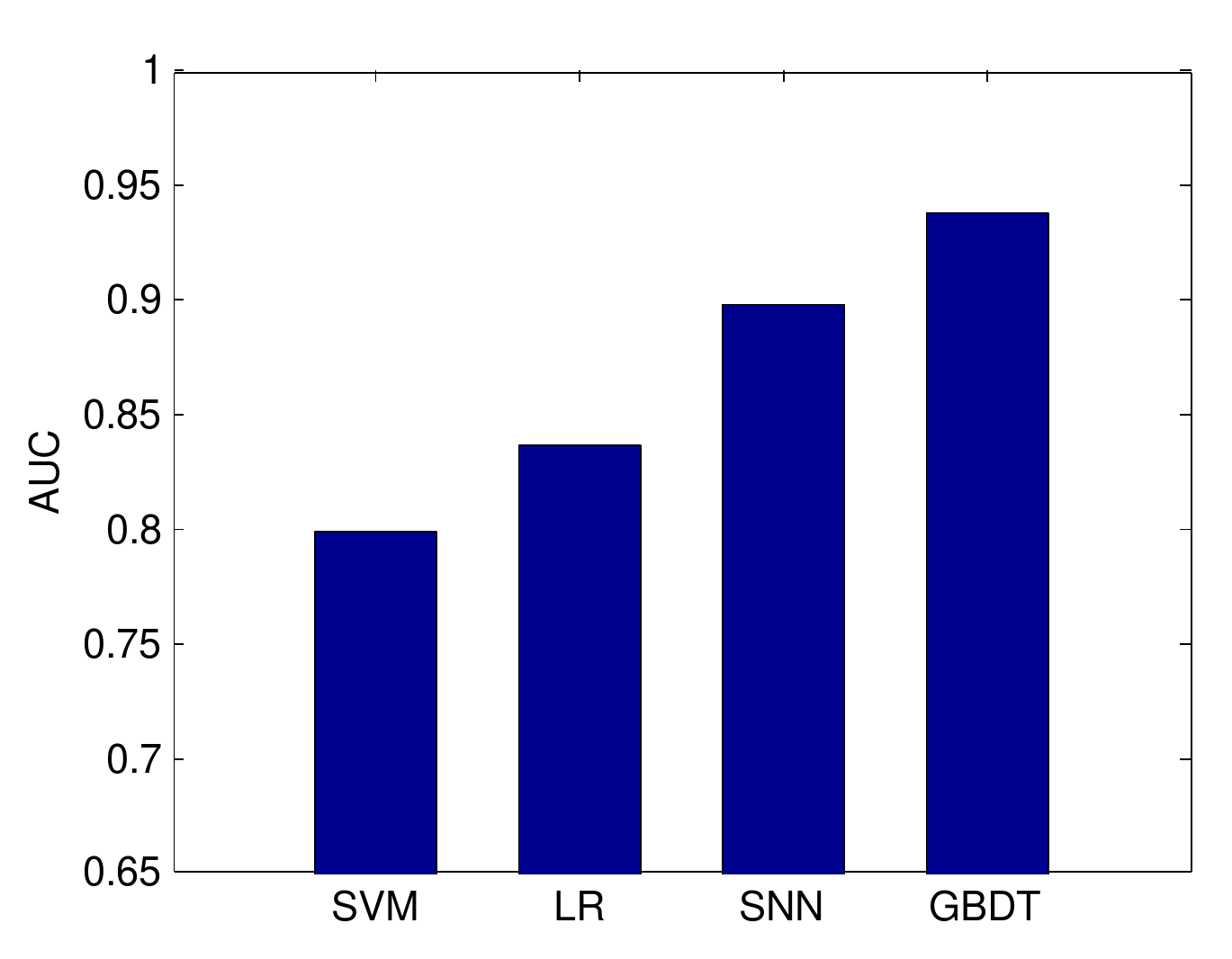}
  \caption{AUC}
  \label{fig:auc}
\end{subfigure} 
\begin{subfigure}{.24\textwidth}
  \centering
  \includegraphics[width=1\textwidth]{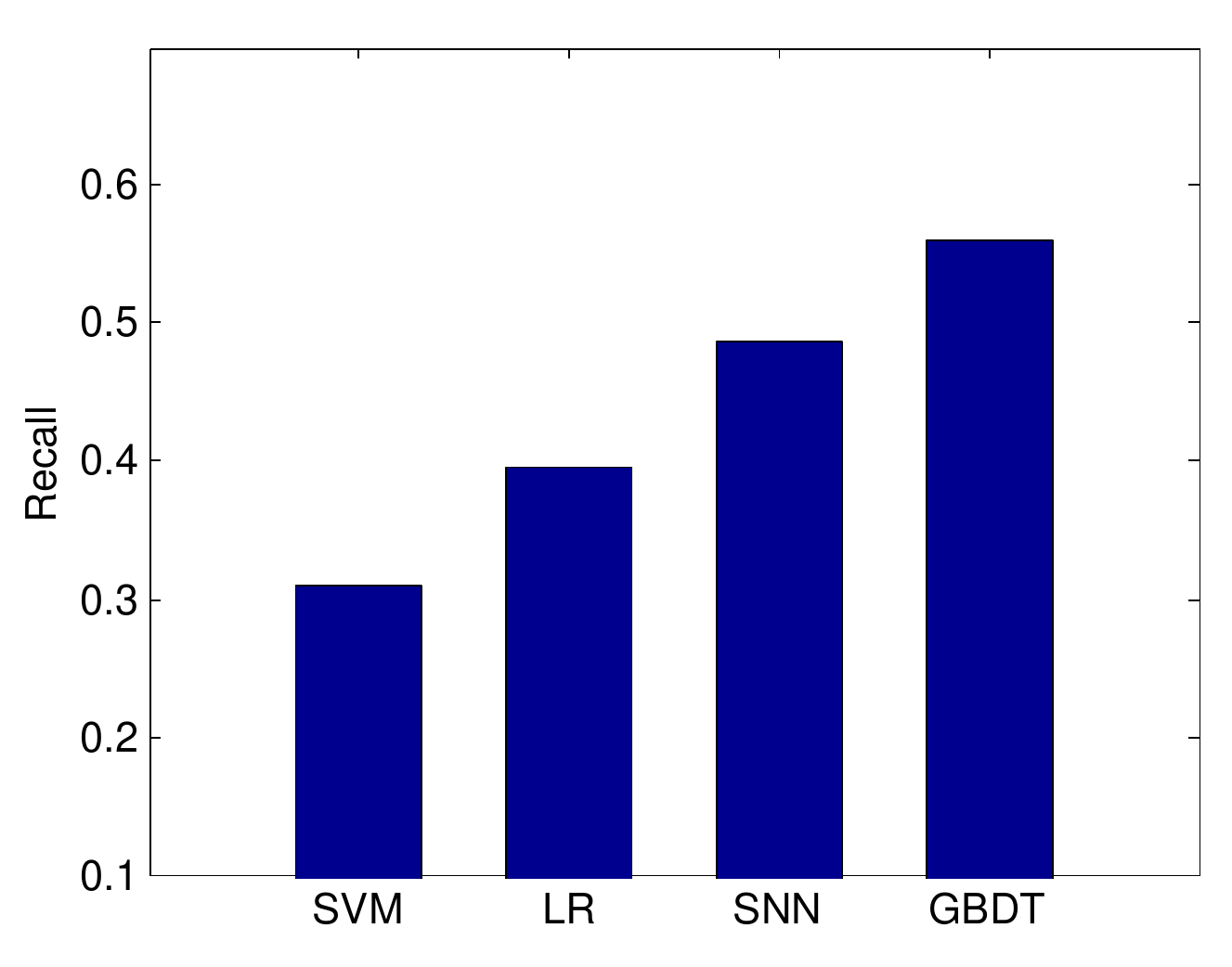}
  \caption{Recall}
  \label{fig:recall}
\end{subfigure} 
\begin{subfigure}{.24\textwidth}
  \centering
  \includegraphics[width=1\textwidth]{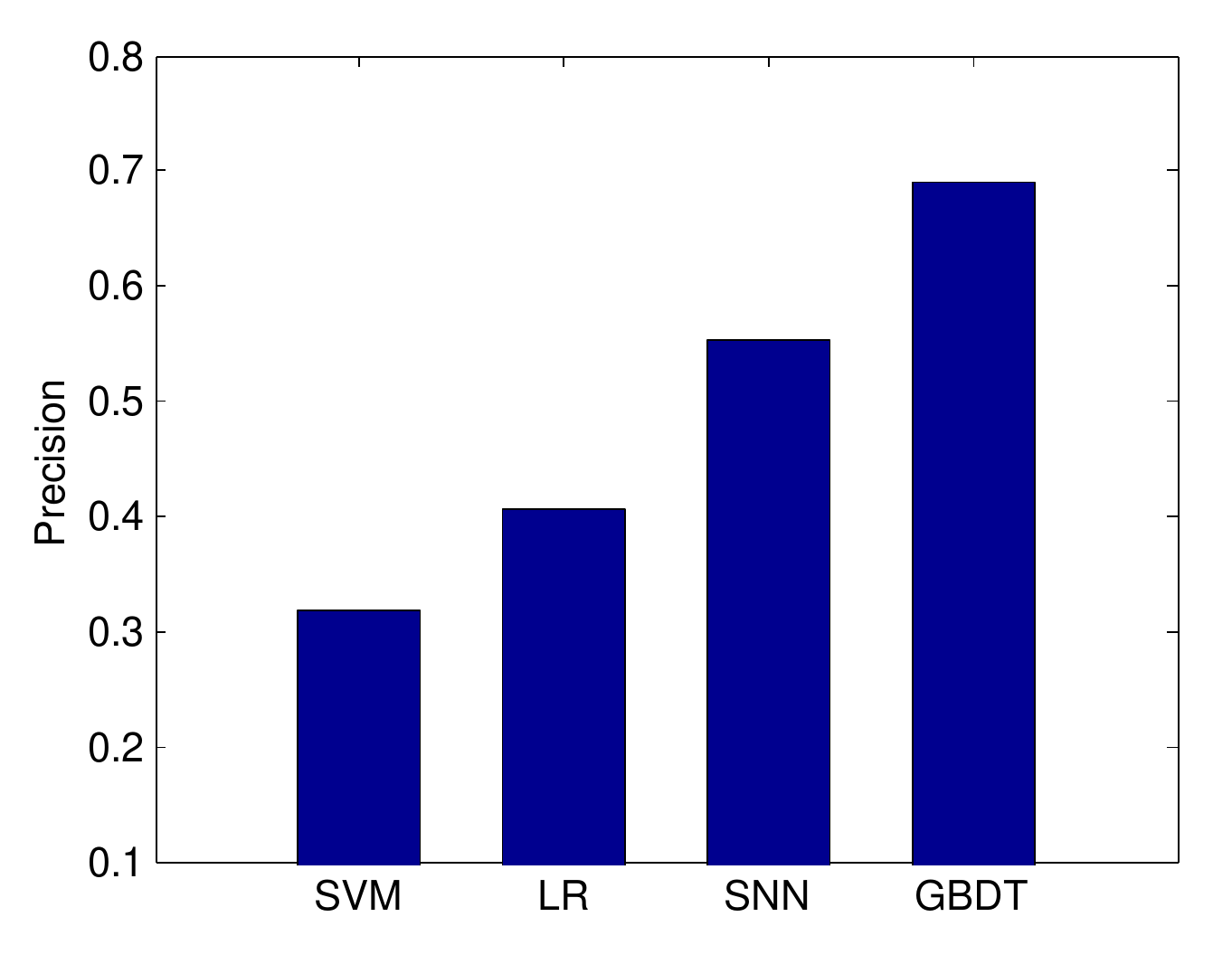}
  \caption{Precision}
  \label{fig:precision}
\end{subfigure} 
\begin{subfigure}{.24\textwidth}
  \centering
  \includegraphics[width=1\textwidth]{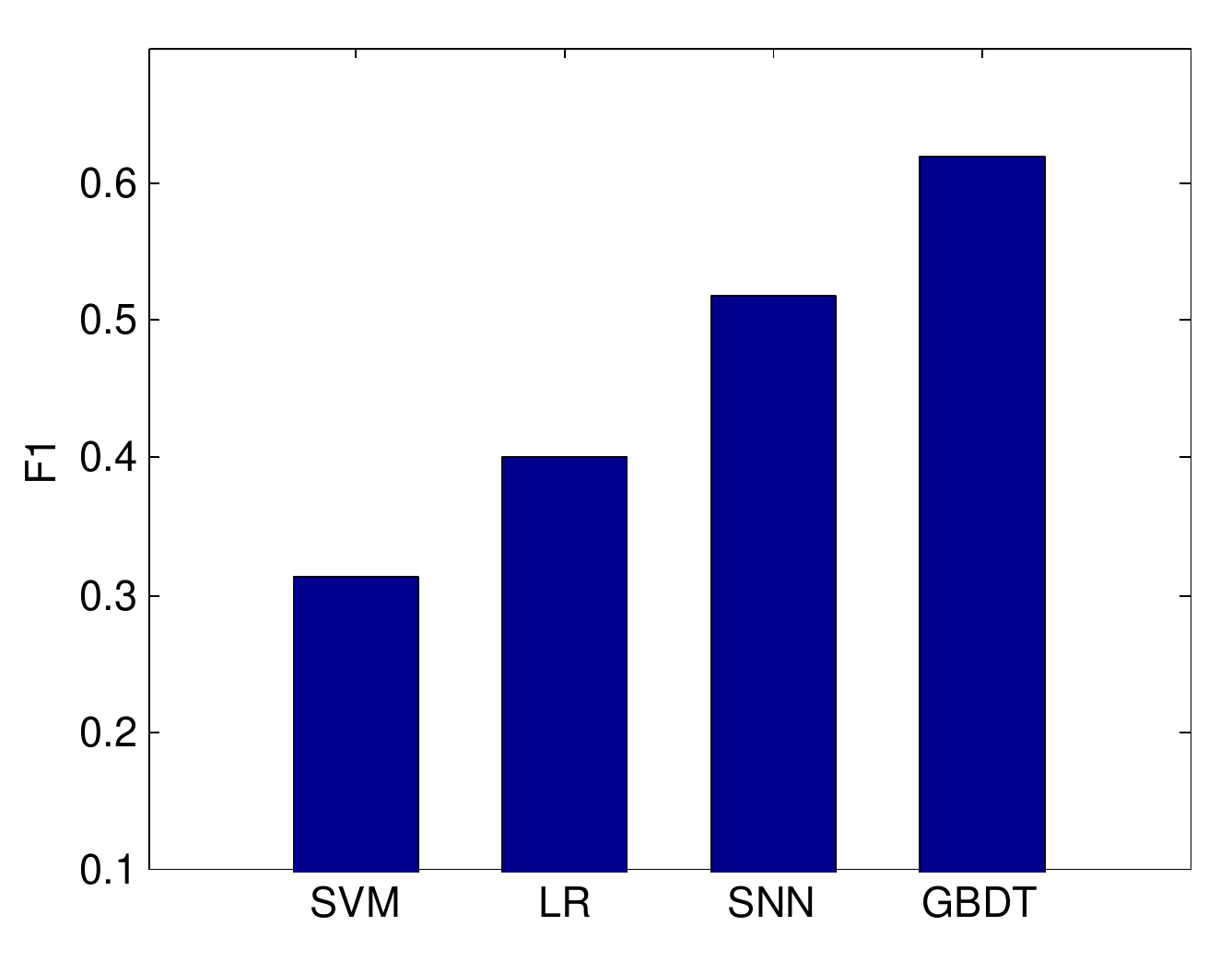}
  \caption{F1}
  \label{fig:F1}
\end{subfigure} 
\caption{Performance comparison among SVM, Logistic Regression, Shallow Neural Network, and Gradient Boosted Decision Tree.}
\label{fig:perfcomp_model}
\end{figure*}

\section{Negative Sampling}
There are a total of 339,405 unique users in the browse log, where the number of users in the training set is 240,732. As described in the last section, for each instance (user-user pair), we extract 621 features. We randomly output 10,000 instances and found that the file size is about 17.74MB. In this way, enumerating the whole user-user pairs will result in 57,951,895,824 instances in the training file, which requires about 100TB in space. Thus we have to do negative instances sampling. We denote  $\mathbf{U}$ as the user set, $\mathbf{M}$ as the matching pairs, $\mathbf{S}$ to be the sampled training instances. We propose an iterative negative sampling algorithm which is shown in Algorithm \ref{alg:ns}. We use logistic regression as the kernel model for instances selection due to its computational efficiency.

\begin{table}
    \centering
    \caption{Performance evaluation for iterative negative sampling algorithm.}
    \begin{tabular}{c|c|c|c|c} \hline \hline
  Sampling method  & AUC  & Recall  & Precision  & F1  \\ \hline
  Random &\ 0.8433  \  & \  0.4195  \   &\  0.4301   \  &\  0.4247   \    \\ \hline 
  INS  & \textbf{0.9383}  & \textbf{0.5613}  & \textbf{0.6906}  & \textbf{0.6193}  \\ \hline \hline
    \end{tabular} 
    \label{tab:nsperf}
\end{table}

In Algorithm \ref{alg:ns}, $n$ is usually small, in our case we set $n=10$. There are some tricks to selecting top $\frac{n}{2}$ users for each specific user $u$. We don't need to go through all the 240,732 users. E.g., we can randomly pick up 1000 users and select the top $\frac{n}{2}$ users for $u$. Then we iterate for 10 times. During the competition we put enough patient to compute all the 240,732 users for each user $u$. We run the program in parallel on 9 machines and one iteration costs about 20 hours. In this condition, one iteration is enough to achieve good performance. We compare the performance of randomly sampling and our proposed negative sampling algorithm, the results are shown in Table \ref{tab:nsperf}.

\section{Model Selection}
We compare the performance of several models, including Gradient Boosted Decision Tree (GBDT) \cite{friedman2001greedy}, Logistic Regression (LR), Shallow Neural Network (SNN), and SVM (with linear kernel). Figure \ref{fig:perfcomp_model} showns the results, from which we can observe that GBDT significantly outperforms the other three models. It is in accordance with expectation because gradient boosting machine is the state-of-the-art classification model as reported from most of data mining competitions\footnote{https://www.kaggle.com/wiki/PastSolutions}.

A golden rule for achieving the top rank in data mining competitions is that we need to train various models and make ensembles \cite{Liu:2016:RBP:2939672.2939674,jianxun}. However, it is usually time consuming to find an optimal way to ensemble. We didn't join the competition until the last week of the end of the competition. So we propose a simple but turns out to be efficient way to ensemble: from the training data we train two model, LR and GBDT. We use the LR model to select 100 candidates for each user in the test set. After this we can get a smaller test set. Then we use the GBDT model to make predictions on the smaller test set. Table \ref{tab:LRfiltertest} shows that this simple approach greatly improves the accuracy of the prediction. One possible explanation is that there are many negative instances which non-linear (tree) model such as GBDT could not differentiate from positive instances. However, linear models like logistic regression happens to work well on this part. Being aware of this, the results shown in  Figure \ref{fig:perfcomp_model} are unfair to LR because for the test set we already apply LR to select top 100 candidates in order to reduce the test space. Due to time limits we did not spend more effort on the model ensemble. However, the aforementioned result implies that there are still opportunities to make improvements.

\begin{table}
    \centering
    \caption{Performance of GBDT without and with LR filtering.}
    \begin{tabular}{c|c|c|c|c} \hline \hline  
  Model  & AUC  & Recall  & Precision  & F1  \\ \hline
  No LR filtering &\ 0.7864  \  & \  0.3551  \   &\  0.4046   \  &\  0.3782   \    \\ \hline 
  With LR filtering  & \textbf{0.9383}  & \textbf{0.5613}  & \textbf{0.6906}  & \textbf{0.6193}  \\ \hline \hline
    \end{tabular} 
    \label{tab:LRfiltertest}
\end{table}

\begin{table}
    \centering
    \caption{Top 5 teams on final leader board.}
    \begin{tabular}{|c|c|c|c|} \hline  
  Rank    & F1  & Precision  & Recall  \\ \hline
  1   & \  0.42038  \   &\  0.39875   \  &\  0.44449   \    \\ \hline 
  2   & \  0.41669  \   &\  0.39444   \  &\  0.44160   \    \\ \hline 
  3   & \  0.41370  \   &\  0.40042   \  &\  0.42790   \    \\ \hline 
  4   & \  0.40168  \   &\  0.36591   \  &\  0.44520   \    \\ \hline 
  5   & \  0.36110  \   &\  0.33227   \  &\  0.39540  \    \\ \hline  
    \end{tabular} 
    \label{tab:finalresult}
\end{table}

\begin{algorithm}  
    \caption{Select Instances for Submission}
    \label{alg:submision}
    \begin{algorithmic}
        \Require predicted test pairs $\mathbf{T}$, and parameters $n, k, r$.
        \Ensure  submission set $\mathbf{P}$  
        %\State
        \State $\mathbf{P} \gets \varnothing $
        \State Add top $n$ pairs from $\mathbf{T}$ to $\mathbf{P}$  
        \For{each user $u$}
            \For{$i = 0 \to k-1$}
                \State retrieve the i-th top predicted user $v$ for $u$
                \If{ $Ranking( \mathbf{T}, \langle u,v\rangle )  < n \times r$}
                    \State Add $\langle u,v\rangle$ to $\mathbf{P}$ if not exists
                \EndIf
            \EndFor
        \EndFor
        \State \Return{$\mathbf{P}$}
    \end{algorithmic}
\end{algorithm}

\section{Online Evaluation}
Every time we make improvements to local evaluations, the corresponding online F1-score also improves. It indicates that our framework does not cause overfitting and the local test set is extracted  appropriately. There are a total of 215,307 true pairs in the test set. One small trick is that since F1-score is a trade-off between recall and precision, we don't need to submit too many predicted instances in order to achieve a peak F1 score. We only included about 100,000 instances in our final submission file. The final post processing algorithm is shown in Algorithm \ref{alg:submision}. Besides top $n$ pairs, for each user in the test set we also select the top $k$ candidates whose global rank is not far-away from $n$. We ended up with 2nd place on the leader board. Table \ref{tab:finalresult} lists the top 5 teams' scores. The top 3 teams' final scores are very close.

\section{conclusion}
In this paper, we describe our solution for the \textsl{CIKM Cup 2016 User Linking Challenge} at which we took the second place of the competition. It has three primary componnets that we focus on: the feature engineering, negative sampling, and model selection. Since time was limited when our approach was conceived, there are still many possible approaches which we have not tried yet. For example, learning-to-rank \cite{burges2010ranknet} is a promising approach to this competition; and we can also apply other ensemble methods such as stacking.

\section{Related work}
The topic of CIKM Cup 2016 (track 1) is very similar to the ICDM Cup 2015\footnote{https://www.kaggle.com/c/icdm-2015-drawbridge-cross-device-connections}:\textsl{Drawbridge Cross-Device Connections}, except that the two events provide different types of data for mining. Among the winning solutions\cite{walthers2015learning,Kejela:2015:CCI:2919331.2919558,diaz2015cross}, learning-to-rank and binary classification are the two most popular paradigms. \cite{walthers2015learning} points out that learning-to-rank is more suitable for cross-device linking problem because for each entity, we don't need the absolute value of its probability in matching with another entity, instead what we need is the relative ranking according the target entity. \cite{diaz2015cross} combines several techniques, such as semi-surpervised learning and bagging, to further boost the performance. Since it is not feasible to generate a full entity-to-entity pair set, down-sampling is used by all the winning solutions. However, they down-sample the candidates by some particular rules. In this paper, we propose a negative sampling method which selects candidates iteratively with a weak learner.

Cross-device user matching is also related to link prediction \cite{liben2007link,al2006link,yin2010unified}. From the graph-thoretical perspective, users can be regarded as nodes and a user-matching can be modeled as an edge between the two corresponding nodes. Thus the task is to predict the missing links in the graph. \cite{getoor2005link} surveys several well studied link mining tasks and methods. In this paper, we model the task as a binary classification problem for simplicity. In the next steps we will study how to make breakthroughs using graph models.

\bibliographystyle{abbrv}
\bibliography{myref} 

\begin{thebibliography}{10}

\bibitem{al2006link}
M.~Al~Hasan, V.~Chaoji, S.~Salem, and M.~Zaki.
\newblock Link prediction using supervised learning.
\newblock In {\em SDM06: workshop on link analysis, counter-terrorism and
  security}, 2006.

\bibitem{burges2010ranknet}
C.~J. Burges.
\newblock From ranknet to lambdarank to lambdamart: An overview.
\newblock Technical report.

\bibitem{diaz2015cross}
R.~D{\'\i}az-Morales.
\newblock Cross-device tracking: Matching devices and cookies.
\newblock {\em arXiv preprint arXiv:1510.01175}, 2015.

\bibitem{friedman2001greedy}
J.~H. Friedman.
\newblock Greedy function approximation: a gradient boosting machine.
\newblock {\em Annals of statistics}, pages 1189--1232, 2001.

\bibitem{getoor2005link}
L.~Getoor and C.~P. Diehl.
\newblock Link mining: a survey.
\newblock {\em ACM SIGKDD Explorations Newsletter}, 7(2):3--12, 2005.

\bibitem{Kejela:2015:CCI:2919331.2919558}
G.~Kejela and C.~Rong.
\newblock Cross-device consumer identification.
\newblock In {\em Proceedings of the 2015 IEEE International Conference on Data
  Mining Workshop (ICDMW)}, ICDMW '15, pages 1687--1689, Washington, DC, USA,
  2015. IEEE Computer Society.

\bibitem{jianxun}
J.~Lian, X.~Xie, and G.~Sun.
\newblock {Winning the Second Place of IJCAI-15 Repeat Buyers Prediction
  Contest: a Feature Engineering Approach}.
\newblock
  \url{https://github.com/Leavingseason/Competitions/blob/master/Jianxun_IJCAI_Cup2015.pdf},
  2015.

\bibitem{liben2007link}
D.~Liben-Nowell and J.~Kleinberg.
\newblock The link-prediction problem for social networks.
\newblock {\em Journal of the American society for information science and
  technology}, 58(7):1019--1031, 2007.

\bibitem{Liu:2016:RBP:2939672.2939674}
G.~Liu, T.~T. Nguyen, G.~Zhao, W.~Zha, J.~Yang, J.~Cao, M.~Wu, P.~Zhao, and
  W.~Chen.
\newblock Repeat buyer prediction for e-commerce.
\newblock In {\em Proceedings of the 22Nd ACM SIGKDD International Conference
  on Knowledge Discovery and Data Mining}, KDD '16, pages 155--164, New York,
  NY, USA, 2016. ACM.

\bibitem{walthers2015learning}
J.~Walthers.
\newblock Learning to rank for cross-device identification.
\newblock In {\em 2015 IEEE International Conference on Data Mining Workshop
  (ICDMW)}, pages 1710--1712. IEEE, 2015.

\bibitem{yin2010unified}
Z.~Yin, M.~Gupta, T.~Weninger, and J.~Han.
\newblock A unified framework for link recommendation using random walks.
\newblock In {\em Advances in Social Networks Analysis and Mining (ASONAM),
  2010 International Conference on}, pages 152--159. IEEE, 2010.

\end{thebibliography}
\end{document}